# Use of Eigenvector Centrality to Detect Graph Isomorphism

Dr. Natarajan Meghanathan
Professor of Computer Science
Jackson State University
Jackson MS 39217, USA
E-mail: natarajan.meghanathan@jsums.edu

**Abstract**

Graph Isomorphism is one of the classical problems of graph theory for which no deterministic polynomial-time algorithm is currently known, but has been neither proven to be NP-complete. Several heuristic algorithms have been proposed to determine whether or not two graphs are isomorphic (i.e., structurally the same). In this research, we propose to use the sequence (either the non-decreasing or non-increasing order) of eigenvector centrality (EVC) values of the vertices of two graphs as a precursor step to decide whether or not to further conduct tests for graph isomorphism. The eigenvector centrality of a vertex in a graph is a measure of the degree of the vertex as well as the degrees of its neighbors. We hypothesize that if the non-increasing (or non-decreasing) order of listings of the EVC values of the vertices of two test graphs are not the same, then the two graphs are not isomorphic. If two test graphs have an identical non-increasing order of the EVC sequence, then they are declared to be potentially isomorphic and confirmed through additional heuristics. We test our hypothesis on random graphs (generated according to the Erdos-Renyi model) and we observe the hypothesis to be indeed true: graph pairs that have the same sequence of non-increasing order of EVC values have been confirmed to be isomorphic using the well-known Nauty software.

## 1 Introduction

Graph isomorphism is one of the classical problems of graph theory for which there exist no deterministic polynomial-time algorithm and at the same time the problem has not been yet proven to be NP-complete. Given two graphs $G_1(V_1, E_1)$ and $G_2(V_2, E_2)$ - where $V_1$ and $E_1$ are the sets of vertices and edges of $G_1$ and $V_2$ and $E_2$ are the sets of vertices and edges of $G_2$ - we say the two graphs are isomorphic, if the two graphs are structurally the same. In other words, two graphs $G_1(V_1, E_1)$ and $G_2(V_2, E_2)$ are isomorphic [1] if and only if we can find a bijective mapping $f$ of the vertices of $G_1$ and $G_2$, such that $\forall v \in V_1, f(v) \in V_2$ and $\forall (u, v) \in E_1, (f(u), f(v)) \in E_2$. As the problem belongs to the class NP, several heuristics (e.g., [7-9]) have been proposed to determine whether any two graphs $G_1$ and $G_2$ are isomorphic or not. The bane of these heuristics is that they are too time-consuming for large graphs and could lead to identifying several false positives (i.e., concluding a pair of two non-isomorphic graphs as isomorphic).

To minimize the computation time, the test graphs (graphs that are to be tested for isomorphism) are subject to one or more precursor steps (pre-processing routines) that could categorically discard certain pair of graphs as non-isomorphic (without the need for validating further using any time-consuming heuristic). For two graphs $G_1(V_1, E_1)$ and $G_2(V_2, E_2)$ to be isomorphic, a basic requirement is that the two graphs should have the same number of vertices and similarly the same number of edges. That is, if $G_1(V_1, E_1)$ and $G_2(V_2, E_2)$ are to be isomorphic, then it implies $|V_1| = |V_2|$ and $|E_1| = |E_2|$. If $|V_1| \neq |V_2|$ and/or $|E_1| \neq |E_2|$, then we can categorically say that $G_1$ and $G_2$ are not isomorphic and the two graphs need not be processed further through any time-consuming heuristics to test for isomorphism.

In addition to checking for the number of vertices and edges, one of the common precursor steps to test for graph isomorphism is to determine the degree of the vertices of the two graphs that are to be tested for isomorphism and check if a non-increasing order (or a non-decreasing order; we will follow a convention of sorting in a non-increasing order) of the degrees of the vertices of the two graphs is the same. If the non-increasing order of the degree sequence of two graphs $G_1$ and $G_2$ are not the same, then the two graphs can be categorically ruled out from being isomorphic. If two graphs are isomorphic, then identical degree sequence of the vertices in a particular sorted order is a necessity. However as shown in Figure 1, it is possible that two graphs could have the same degree sequence in a particular sorted order, but need not be isomorphic [2]. Though very time-efficient, the degree sequence-based precursor step to test for graph isomorphism is typically considered to be erratic and not reliable (leading to false positives), especially while testing for isomorphism among graphs with a smaller number of vertices (like the example in Figure 1).

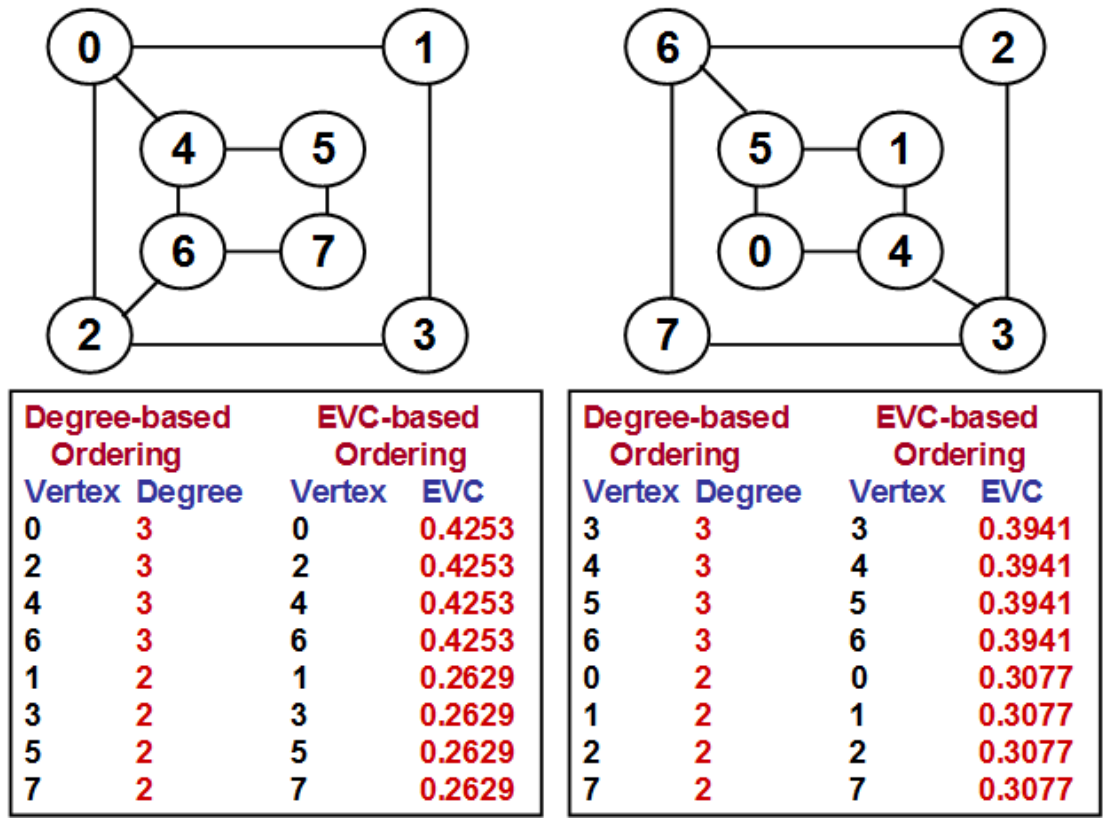

**Figure 1:** Example for Two Non-Isomorphic Graphs with the Same Degree Sequence, but Different Eigenvector Centrality (EVC) Sequence

Centrality metrics are one of the commonly used quantitative measures to rank the vertices of a graph based on the topological structure of the graph [3]. Degree centrality is one of the primitive and typically used centrality metrics for complex network analysis; but, in addition to the weakness illustrated in Figure 1 and explained in the previous paragraph, it is also evident from Figure 1 that degree centrality-based ranking of the vertices could result in ties (i.e., the technique has weak discrimination power) among vertices having the same degree (as the degree centrality values are integers) and it may not be possible to unambiguously rank the vertices; for graphs of any size, it is likely that more than one vertex may have the same degree (ties). Eigenvector centrality (EVC) is a well-known centrality measure in the area of complex networks [4]. The EVC of a vertex is a measure of the degree of the vertex as well as the degree of its neighbors (calculations of EVC values is discussed in Section 2). For example: if two vertices $X$ and $Y$ have degree 3, but if all the three neighbors of $X$ have a degree 2 and if at least one of the neighbors of $Y$ have degree greater than 2 and others have degree at least 2, then the EVC of $Y$ is guaranteed to be greater than the EVC of $X$. In general, the EVC of a vertex not only depends on the degree of the vertex, but also on the degree of its neighbors. For a connected graph, the EVC values of the vertices are positive real numbers in the range (0...1) and are more likely to be different from each other, contributing to the scenario of unambiguous ranking of the vertices as much as possible (the EVC technique has a relatively stronger discrimination power compared to the degree-based technique).

With respect to Figure 1, we notice that the non-increasing order listings of the EVC values of the vertices for the two graphs are not the same. The discrepancy is obvious in the largest EVC value of the two sequences itself. The largest EVC value for a vertex in the first graph is 0.4253 and the largest EVC value for a vertex in the second graph is 0.3941. The example in Figure 1 is a motivation for our hypothesis to use the EVC values as the basis for deciding whether or not two graphs could be isomorphic.

The rest of the paper is organized as follows: Section 2 explains the procedure to determine the Eigenvector Centrality (EVC) values of the vertices. In Section 3, we propose the use of the Eigenvector Centrality (EVC) measure as the basis of the precursor step to determine whether or not two graphs are isomorphic. In Section 4, we test our hypothesis on random network graphs (generated according to the Erdos-Renyi model [5]) with regards to the application of the EVC measure for detecting isomorphism among graphs. Section 5 discusses related work. Section 6 concludes the paper. Throughout the paper, the terms 'node' and 'vertex' as well as 'edge' and 'link' are used interchangeably. They mean the same.

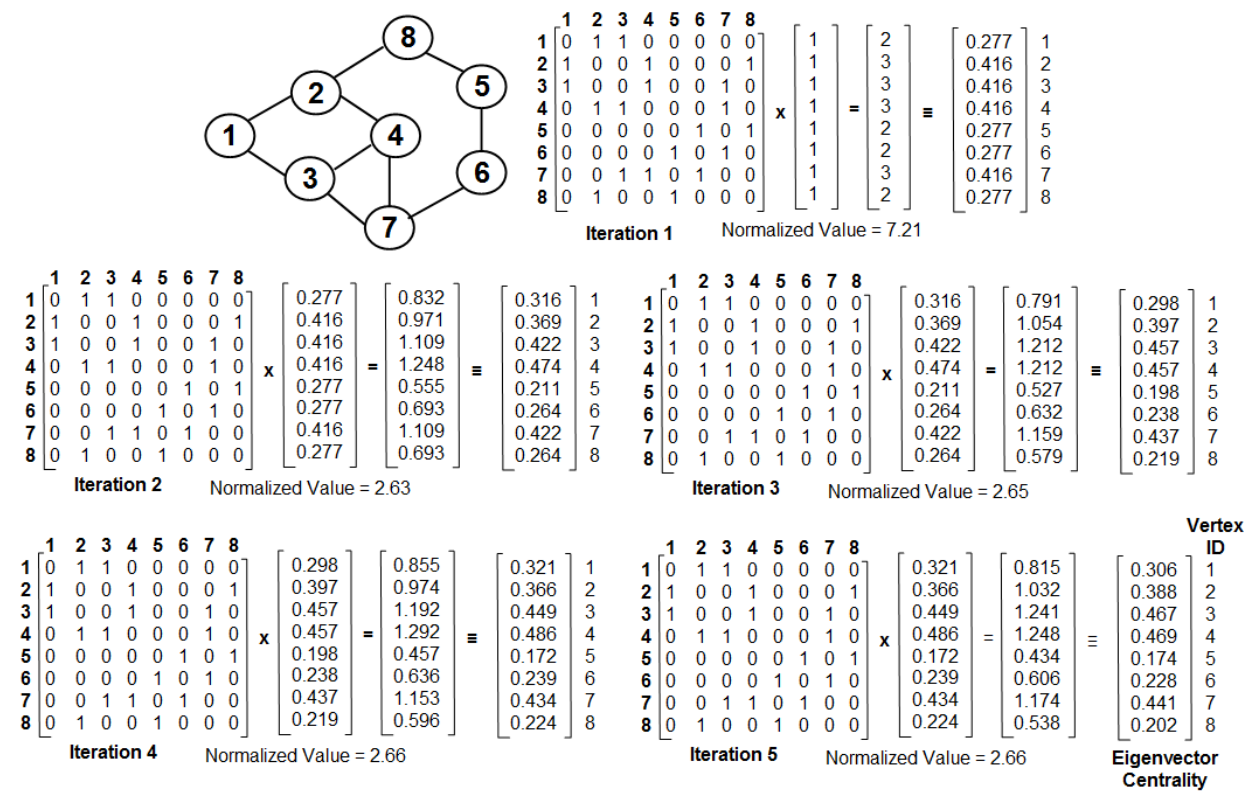

**Figure 2:** Example to Illustrate the Computation of Eigenvector Centrality (EVC) of the Vertices using the Power-Iteration Method

## 2 Eigenvector Centrality (EVC)

The Eigenvector Centrality (EVC) of a vertex is a measure of the degree of the vertex as well as the degree of its neighbors. The EVC of the vertices in a network graph is the principal eigenvector of the adjacency matrix of the graph. The principal eigenvector has an entry for each of the $n$-vertices of the graph. The larger the value of this entry for a vertex, the higher is its ranking with respect to EVC. We illustrate the use of the Power-iteration method [6] (see example in Figure 2) to efficiently calculate the principal eigenvector for the adjacency matrix of a graph. The eigenvector $X_{i+1}$ of a network graph at the end of the $(i+1)^{th}$ iteration is given by: $X_{i+1} = \dfrac{AX_i}{\|AX_i\|}$, where $\|AX_i\|$ is the normalized value of the product of the adjacency matrix A of a given graph and the tentative eigenvector $X_i$ at the end of iteration $i$. The initial value of $X_i$ is the transpose of [1, 1, ..., 1], a column vector of all 1s, where the number of 1s correspond to the number of vertices in the graph. We continue the iterations until the normalized value $\|AX_{i+1}\|$ converges to that of the normalized value $\|AX_i\|$. The value of the column vector $X_i$ at this juncture is declared the Eigenvector centrality of the graph; the entries corresponding to the individual rows in $X_i$ represent the Eigenvector centrality of the vertices of the graph. The converged normalized value of the Eigenvector is referred to as the Spectral radius.

As can be seen in the example of Figure 2, the EVC of a vertex is a function of both its degree as well as the degree of its neighbors. For instance, we see that both vertices 2 and 4 have the same degree (3); however, vertex 4 is connected to three vertices that have a high degree (3); whereas vertex 2 is connected to two vertices that have a relatively low degree (of degree 2); hence, the EVC of vertex 4 is larger than that of vertex 2. As can be seen in the example of Figure 2, the EVC values of the vertices are more likely to be distinct and could be a better measure for unambiguously ranking the vertices of a network graph.

The number of iterations needed for the normalized value of the eigenvector to converge is anticipated to be less than or equal to the number of vertices in the graph [6]. Each iteration of the power-iteration method requires $\Theta(V^2)$ multiplications, where $V$ is the number of vertices in the graph. With a maximum of $V$ iterations expected, the overall time complexity of the algorithm to determine the Eigenvector Centrality of the vertices of a graph of $V$ vertices is $\Theta(V^3)$.

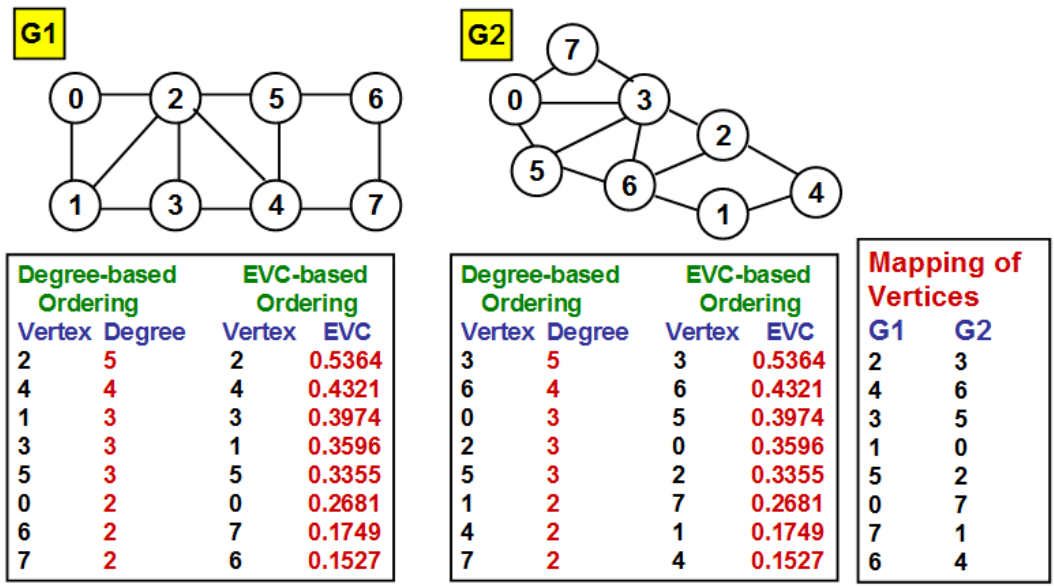

**Figure 3:** Illustration of the Hypothesis: Eigenvector Centrality (EVC) to Decide Graph Isomorphism

## 3 Hypothesis

Our hypothesis is that if a non-increasing order of listings of the EVC values of the vertices for two graphs $G_1$ and $G_2$ are not identical, then the two graphs are not isomorphic. If the non-increasing sequence of EVC values for the two graphs is identical, we declare the two graphs to be potentially isomorphic and subject them to further tests for isomorphism (for confirmation). Thus, the technique of listing the EVC sequence of the vertices (in a non-increasing order) could be used as an effective precursor step before subjecting the graphs to any time-consuming heuristic for graph isomorphism. As the EVC values of the vertices in any random graph are more likely to be unique, this test would also help us to extract a mapping of the vertices between two graphs that have been identified to be potentially isomorphic and make it more easy for the time-consuming complex heuristics to test for isomorphism. We illustrate our hypothesis using an example in Figure 3. From the example, it is very obvious that if two graphs have an identical non-increasing order listing of the EVC sequence, they should have identical non-increasing order listing of the degree sequence; but, not vice-versa (refer example in Figure 1). If two graphs have a different non-increasing order of degree sequence, they cannot have the same non-increasing order of EVC sequence and we do not need to compute the EVC values.

We notice from Figure 3 that the vertices corresponding to the non-increasing order of the EVC values in both the graphs could be uniquely mapped to each other on a one-to-one basis (bijective mapping). On the other hand, the non-increasing order of the degree sequence of the vertices merely facilitates us to group the vertices into different equivalence classes (all vertices of the same degree in both the graphs are said to be equivalent to each other); but, one could not arrive at a unique one-to-one mapping of the vertices that corresponds to the structure of the two graphs. We thus hypothesize that the EVC approach could not only help us to determine whether or not two graphs are isomorphic, it also facilitates us to potentially arrive at a unique one-to-one mapping of the vertices in the corresponding two graphs and feed such a mapping as input to any heuristic that is used to confirm whether two graphs that have been identified to be possibly isomorphic (using the EVC approach) are indeed isomorphic.

## 4  Simulations

We tested our hypothesis by conducting extensive simulations on random network graphs generated according to the Erdos-Renyi model [5]. According to this model, the network has $N$ nodes and the probability of a link between any two nodes is $p_{link}$. For any pair of vertices $u$ and $v$, we generate a random number in the range [0...1] and if the random number is less than $p_{link}$, there is a link between the two vertices $u$ and $v$; otherwise, not. We constructed random networks of $N = 10$ nodes with $p_{link}$ values of 0.2 to 0.8 (in increments of 0.1). We constructed a suite of 1000 networks for each value of $p_{link}$. We chose a smaller value for the number of nodes as we did not observe any pair of isomorphic graphs in a suite of 1000 graphs created with $N = 100$ nodes for any $p_{link}$ value. Even for networks of $N = 10$ nodes, there is a high chance of observing pairs of isomorphic graphs only under low or high values of $p_{link}$. For $p_{link}$ values of 0.2 and 0.3, the pairs of isomorphic graphs observed were typically trees (graphs without any cycles) that have the minimal number of edges to keep all the nodes connected. As we increase the number of links in the networks, the chances of finding any two distinct isomorphic random graphs get extremely small. On the other hand, for $p_{link}$ values of 0.7 and 0.8, the isomorphic graphs were observed to be close to complete graphs (with only one or two missing links per node from becoming a complete graph).

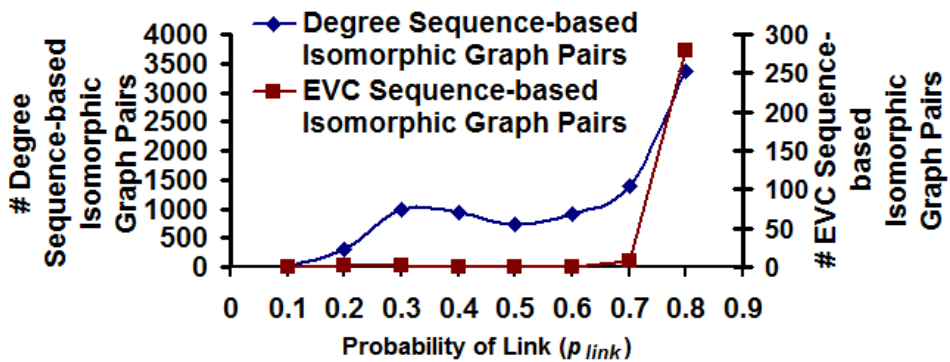

**Figure 4:** Number of Isomorphic Random Graph Pairs: Degree Sequence vs. EVC Sequence Approach

The success of the hypothesis is evaluated by determining the number of pairs of isomorphic graphs identified based on the non-increasing order of the EVC sequence vis-a-vis the degree sequence. As mentioned earlier, if two graphs are isomorphic, then the non-increasing order of listing of the EVC values of the vertices has to be identical (as the two graphs are essentially the same, with just the vertices labeled differently). This implies that if the non-increasing order of listing of the EVC values of the vertices for a pair of graphs $G_1$ and $G_2$ are not identical, we need not further subject the two graphs to any other heuristic test for isomorphism. If two graphs are identified to be potentially isomorphic based on the EVC sequence, we further processed those two graphs using the Nauty software [7] and confirmed that the two graphs are indeed isomorphic to each other. We did not observe any false positives with the EVC approach. The Nauty software [7] is the world's fastest testing software (available at: http://www3.cs.stonybrook.edu/~algorith/implement/nauty/implement.shtml) for graph isomorphism.

Figure 4 illustrates the number of graph pairs that have been identified to be potentially isomorphic on the basis of the EVC sequence approach vis-a-vis the degree sequence approach. We observe that even with the degree sequence approach, for moderate $p_{link}$ values (0.4-0.5), the number of graph pairs identified to be potentially isomorphic decreases from that observed for low-moderate $p_{link}$ value of 0.3. As we further increase the $p_{link}$ value, the number of graph pairs identified to be potentially isomorphic increases significantly with both the degree sequence and EVC sequence-based approach, and the EVC sequence-based approach identifies a significantly larger number of these graph pairs (that are already identified to be potentially isomorphic based on the degree sequence) to be indeed potentially isomorphic and this is further reconfirmed through the Nauty software. For low-moderate $p_{link}$ values, we observe the degree sequence-based approach to identify an increasingly larger number of graph pairs to be potentially isomorphic, but they were observed to be indeed not isomorphic on the basis of the EVC sequence approach as well as when tested using the Nauty software. This vindicates our earlier assertion (in Section 1) that the degree sequence-based precursor step is prone to incurring a larger number of false positives (i.e., erratically identifying graph pairs as isomorphic when they are indeed not isomorphic).

## 5   Related Work

Though centrality measures have been widely used for problems related to complex network analysis [3], the degree centrality measure is the only common and most directly used centrality measure to test for graph isomorphism [1]. The other commonly used centrality-based precursor step to test for the isomorphism of two or more graphs is to find the shortest path vector for each vertex in the test graphs and evaluate the similarity of the shortest path matrix (an ensemble of the shortest path vectors of the constituent vertices) of the test graphs. Since the one-to-one mapping between the vertices of the test graphs is not known a priori, one would need a time-efficient algorithm to compare the columns (shortest path vectors) of two matrices for similarity between the columns. The closeness centrality measure [3] is the centrality measure that matches to the above precursor step. Both the degree and closeness centrality measures have an inherent weakness of incurring only integer values (contributing to their poor discrimination of the vertices) and it is quite possible that two or more vertices have the same integer value under either of these centrality measures and one would not be able to obtain a distinct ranking of the vertices (i.e., unique values of the centrality scores) to detect for graph isomorphism. The eigenvector centrality measure incurs real numbers as values in the range (0...1) and has a much higher chance of incurring distinct values for each of the vertices of a graph. Though there could be scenarios where two or more vertices have the same EVC value, a non-increasing or non-decreasing order listing the EVC values of the vertices of two different graphs is more likely to be different from each other if the two graphs are non-isomorphic. As the complexity of the graph topology increases (as the number of vertices and edges increases), we observed it to be extremely difficult to generate two random graphs that have the same sequence (say in the non-increasing order) of EVC values for the vertices and be isomorphic.

As mentioned earlier, graph isomorphism is one of the classical problems of graph theory that has not been yet proven to be NP-complete, but there does not exist a deterministic polynomial time algorithm either. Many heuristics have been proposed to solve the graph isomorphism problem (e.g., Nauty [7], Ullmann algorithm [8] and VF2 [9]), but all of them take an exponential time at the worst case as most of them take the approach of progressively searching for all possible matching between the vertices of the test graphs. To reduce the search complexity, the heuristics could use precursor steps like checking for identical degree sequence for the vertices of the test graphs. It would be preferable to use precursor steps that contribute to fewer false positives, if not none. This is where our proposed approach of using the eigenvector centrality (EVC) fits the bill. We observe from the simulations that all the graphs identified to be isomorphic (using the EVC approach) are indeed isomorphic. Thus, the EVC sequence-based listing of the vertices could be rather used as an effective precursor step to rule out graph pairs that are guaranteed to be not isomorphic, especially when used with the more recently developed time-efficient heuristics that effectively prune the search space (e.g., the parameterized matching [10] algorithm).

The eigenvector centrality (EVC) measure falls under a broad category of measures called "graph invariants" that have been extensively investigated in discrete mathematics [11-12], structural chemistry [13-14] and computer science [15]. These graph invariants can be classified to be either global (e.g., Randic index [16]) or local (e.g., vertex complexity [17]) as well as be either information-theoretic (statistical quantities) [18-19] or non-information-theoretic indices [20]. With the objective of reducing the run-time complexity of the heuristics for graph isomorphism, weaker but time-efficient precursor tests (measures with poor discrimination power like the degree sequence) were rather commonly used. Sometimes, a suite of such simplistic graph invariants were used [21] and test graphs observed to be potentially isomorphic based on each of these invariants were considered for further analysis with a complex heuristic. The discrimination power of the weaker graph invariants also vary with the type of graphs studied [21]. To the best of our knowledge, the discrimination power of the more complex graph invariants - especially those based on the spectral characteristics of a graph (like that of the Eigenvector Centrality), is yet to be analyzed. Ours is the first effort in this direction.

## 6  Conclusions

The high-level contribution of this paper is the proposal to use the Eigenvector Centrality (EVC) measure to detect isomorphism among two or more graphs. We propose that if the non-increasing order (or non-decreasing order) of listing the EVC values of the vertices of the test graphs are not identical, then the test graphs are not isomorphic and need not be further processed by any time-consuming heuristic to detect graph isomorphism. This implies that if two or more graphs are isomorphic to each other, their EVC values written in the non-increasing order must be identical. We test our hypothesis on a suite of random network graphs generated with different values for the probability of link and observed the EVC approach to be effective: there are no false positives, unlike the degree sequence based approach. The graph pairs that are observed to have an identical EVC sequence are confirmed to be indeed isomorphic using the Nauty graph isomorphism detection software. We also observe it to be extremely difficult to generate isomorphic random graphs under moderate values for the probability of link (0.4-0.6); it is rather relatively more easy to generate isomorphic random graphs that are either trees (created when the probability of link values are low: 0.2-0.3) or close to complete graphs (created when the probability of link values are high: 0.7-0.8).